\documentclass[twoside,twocolumn,9pt]{article}
\usepackage{extsizes}
\usepackage[super,sort&compress,comma]{natbib} 
\usepackage[version=3]{mhchem}
\usepackage[left=1.5cm, right=1.5cm, top=1.785cm, bottom=2.0cm]{geometry}
\usepackage{balance}
\usepackage{times,mathptmx}
\usepackage{sectsty}
\usepackage{graphicx} 
\usepackage{lastpage}
\usepackage[format=plain,justification=justified,singlelinecheck=false,font={stretch=1.125,small,sf},labelfont=bf,labelsep=space]{caption}
\usepackage{float}
\usepackage{fancyhdr}
\usepackage{fnpos}
\usepackage[english]{babel}
\addto{\captionsenglish}{%
  
}
\usepackage{array}
\usepackage{droidsans}
\usepackage{charter}
\usepackage[T1]{fontenc}
\usepackage[usenames,dvipsnames]{xcolor}
\usepackage{setspace}
\usepackage[compact]{titlesec}
\usepackage{hyperref}
\definecolor{cream}{RGB}{222,217,201}

\begin{document}

\pagestyle{fancy}
\thispagestyle{plain}
\fancypagestyle{plain}{}

%%%PAGE SETUP - Please do not change any commands within this section%%%
\makeFNbottom
\makeatletter
\renewcommand\LARGE{\@setfontsize\LARGE{15pt}{17}}
\renewcommand\Large{\@setfontsize\Large{12pt}{14}}
\renewcommand\large{\@setfontsize\large{10pt}{12}}
\renewcommand\footnotesize{\@setfontsize\footnotesize{7pt}{10}}
\makeatother

\renewcommand{\thefootnote}{\fnsymbol{footnote}}
\renewcommand\footnoterule{\vspace*{1pt}% 
\color{cream}\hrule width 3.5in height 0.4pt \color{black}\vspace*{5pt}} 
\setcounter{secnumdepth}{5}

\makeatletter 
\renewcommand\@biblabel[1]{#1}            
\renewcommand\@makefntext[1]% 
{\noindent\makebox[0pt][r]{\@thefnmark\,}#1}
\makeatother 
\renewcommand{\figurename}{\small{Fig.}~}
\sectionfont{\sffamily\Large}
\subsectionfont{\normalsize}
\subsubsectionfont{\bf}
\setstretch{1.125} %In particular, please do not alter this line.
\setlength{\skip\footins}{0.8cm}
\setlength{\footnotesep}{0.25cm}
\setlength{\jot}{10pt}
\titlespacing*{\section}{0pt}{4pt}{4pt}
\titlespacing*{\subsection}{0pt}{15pt}{1pt}
%%%END OF PAGE SETUP%%%

%%%TITLE, AUTHORS AND ABSTRACT%%%
\twocolumn[
  \begin{@twocolumnfalse}
\vspace{0cm}
\sffamily
\begin{tabular}{m{0cm} p{18cm} }

%\includegraphics{head_foot/DOI} & \noindent\LARGE{\textbf{The near and far of a pair of magnetic capillary disks$^\dag$}} 
%\vspace{0.3cm} & \vspace{0.3cm} \\

& \noindent\LARGE{\textbf{The near and far of a pair of magnetic capillary disks}} \\%Article title goes here instead of the text "This is the title"
\vspace{0.3cm} & \vspace{0.3cm} \\

 & \noindent\large{Lyndon Koens,$^{\ast}$\textit{$^{a \dag}$} Wendong Wang,\textit{$^{b}$} Metin Sitti,\textit{$^{b,c}$} and Eric Lauga\textit{$^{a}$}} \\%Author names go here instead of "Full name", etc.

& \noindent\normalsize{Control on microscopic scales  depends critically on our ability to manipulate interactions with different physical fields. The creation of micro-machines therefore requires us to understand how multiple fields, such as surface capillary or electro-magnetic, can be used to produce predictable behaviour. Recently, a spinning micro-raft system  was developed  that exhibited both static and dynamic self-assembly [Wang \textit{et al.} (2017) Sci.~Adv.~\textbf{3}, e1602522]. These rafts employed both capillary and magnetic interactions and, at a critical driving frequency, would suddenly change from stable orbital patterns to static assembled structures. In this paper,  we explain the dynamics of two interacting micro-rafts through a combination of theoretical models and experiments. This is first achieved by identifying the governing physics of the orbital patterns, the assembled structures, and the collapse separately. We find that the orbital patterns are determined by the short range capillary interactions between the disks, while the explanations of the other two behaviours only require the capillary far field. Finally we combine the three models to explain the dynamics of a new micro-raft experiment.} \\%The abstrast goes here instead of the text "The abstract should be..."

\end{tabular}

 \end{@twocolumnfalse} \vspace{0.6cm}

  ]
%%%END OF TITLE, AUTHORS AND ABSTRACT%%%

%%%FONT SETUP - please do not change any commands within this section
\renewcommand*\rmdefault{bch}\normalfont\upshape
\rmfamily
\section*{}
\vspace{-1cm}

%%%FOOTNOTES%%%

\footnotetext{\textit{$^{a}$~Department of Applied Mathematics and Theoretical Physics, University of Cambridge, Wilberforce Road, Cambridge CB3 0WA, United Kingdom}}
\footnotetext{\textit{$^{b}$~Physical Intelligence Department, Max Planck Institute for Intelligent Systems, 70569 Stuttgart, Germany.}}
\footnotetext{\textit{$^{c}$~School of Medicine and Mechanical Engineering Department, Koc University, 34450 Istanbul, Turkey}}
%Please use \dag to cite the ESI in the main text of the article.
%If you article does not have ESI please remove the the \dag symbol from the title and the footnotetext below.
\footnotetext{$\ast$~Email: lyndon.koens@mq.edu.au}
\footnotetext{\dag~Present address: Department of Mathematics and Statistics, Macquarie University, 192 Balaclava Rd, Macquarie Park NSW 2113 Australia.}
%additional addresses can be cited as above using the lower-case letters, c, d, e... If all authors are from the same address, no letter is required

%\footnotetext{\ddag~Additional footnotes to the title and authors can be included \textit{e.g.}\ `Present address:' or `These authors contributed equally to this work' as above using the symbols: \ddag, \textsection, and \P. Please place the appropriate symbol next to the author's name and include a \texttt{\textbackslash footnotetext} entry in the the correct place in the list.}

%%%END OF FOOTNOTES%%%

%%%MAIN TEXT%%%%
\section{Introduction}

\begin{figure*}[t]
\begin{center}
\includegraphics[width=0.9\textwidth]{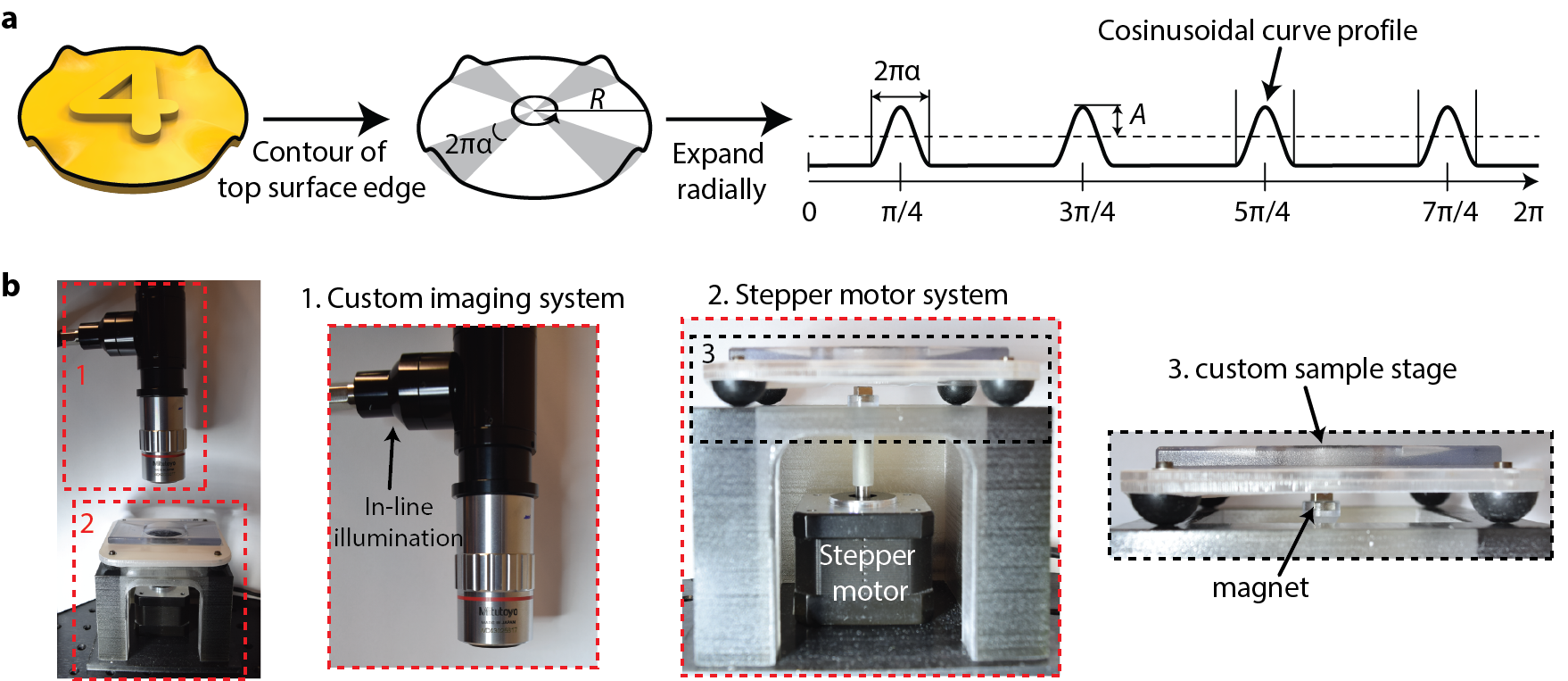}
\caption{The configuration of one representative raft, showing three structural parameters: Amplitude $A$, Radius $R$ and arc angle $2 \pi \alpha$ (a) and the current experimental set-up, showing a custom imaging system and a stepper motor system (b); (a) was adapted from Ref.~\cite{Wang2017} as permitted under the AAAS's License to Publish.}
\label{fig:experiment}
\end{center}
\end{figure*}

The elucidation of the local rules that govern the movement of an individual is at the heart of studying natural collective systems, such as the spontaneous formation of biological membranes \cite{Pomorski2013} or the flocking of fish, ants and birds \cite{Theraulaz686,COUZIN20021,Deutschrsfs20120048}. Inspired by these systems, researchers have constructed macroscopic robot collectives \cite{Rubenstein795,Werfel754} and microscopic colloids \cite{ANIE201005078, Wang2015z, Palacci936, Bricard2013, DaviesWykes2016, Cates2018} to further explore these emergent behaviours. While an individual macroscopic robot can be actuated by on-board power and programmed with microprocessors to respond to its environment, at microscopic scales the design space of the interactions between an individual and its neighbours are constrained by physiochemical principles \cite{Huck2005,Whitesides2002}. As a result, the assembly, actuation, and programmability of microscopic systems require a thorough command of the relevant physiochemical principles \cite{Huck2005}.

The spontaneous organisation of such microscopic many body systems into larger structures is often called self-assembly and is typically classified as either static or dynamic \cite{Huck2005, Whitesides2002}. In static self-assembly the constituents in the final configuration no longer deterministically move \cite{Boker2007, Boles2016}, and so it has been used to explore the formation of stable complex structures \cite{Li2018, Timonen2013, Ke2012, Marras2015, Vella2005, Bowden1999, Lewandowski2010,Botto2012, Yao2013a, Poty2014,Danov2005, Kralchevsky2000}. In dynamic self-assembly, however, the components in the final configuration exhibit deterministic motion, either relative to each other or to a background reference frame \cite{Grzelczak2010, Cates2018}, and so form a proxy to the dynamics of moving individuals \cite{Timonen2013, Bricard2013, Grzybowski2000, Zhang2010, Lushi2014, Wioland2016, Vizsnyiczai2017, Mangal2017, Maggi2015a, Walther2008, Kowalik2016, Wang2015z, Palacci936, DaviesWykes2016}. As most of these microscopic systems exist within viscous fluids, dynamic self-assembly typically requires a continuous input of energy \cite{Grzelczak2010} while static self-assembly does not \cite{Boker2007, Boles2016}. 

The two systems often use different physiochemical principles. For example, static self-assembly has been achieved through DNA origami \cite{Ke2012, Marras2015}, in which the chemical bonding interactions guides the structure formation, and through surface capillary interactions \cite{Bowden1999,Kralchevsky2000,  Vella2005, Danov2005, Lewandowski2010,Botto2012, Yao2013a, ANIE201005078, Poty2014}, in which the surface tension at the air-water interface between the edges of the floating bodies drives the formation of static structures. On the other hand, dynamic self-assembly has been created using external magnetic fields \cite{Grzybowski2000, Zhang2010, Timonen2013, Bricard2013, Koens2016, Montenegro-Johnson2016}, swimming bacteria in confined environments \cite{Lushi2014, Wioland2016, Vizsnyiczai2017}, and phoretic motion due to gradients in chemical composition, temperature or charge \cite{ANIE201005078, Walther2008, Maggi2015a, Montenegro-Johnson2015, Kowalik2016, DaviesWykes2016, Mangal2017}.  

The combination of multiple physiochemical principles, in a single microscopic system, can also create greater complexity and new behaviour. One such pairing is the combination of magnetic fields with capillary interactions \cite{Snezhko2011, Grosjean2015, Lumay2013, Wang2017}. Snezhko and Aranson used these interactions to form colloidal aster structures that sat at an air-water interface and could be deformed by the background field \cite{Snezhko2011}. These deformations  allow  the capture and transport of small floating particles. Similarly, Vandewalle \textit{et al.} created an artificial microswimmer by exposing a collection of floating magnetic spheres to a constant background magnetic field perpendicular to the surface together with an oscillating field parallel to it \cite{Grosjean2015, Lumay2013}. The interaction of the constant background field with the capillary forces caused the spheres to take specific configurations while the parallel field determined the direction of motion.

Recently, the dynamic and static self-assemblage of a collection of spinning magnetic-capillary micro-rafts was also reported \cite{Wang2017}. Small circular disks sat at an air-water interface and were actuated by the rotation of a magnetic background field. Programmability was achieved by varying the edge shape of the rafts in order to induce different capillary interactions. These rafts formed stable orbital patterns at high rotation frequencies. These dynamic configurations depend non-linearly on the driving frequency. At a non-zero critical driving frequency the rafts suddenly collapse to the magnetic centre and assemble into `static' structures related to the edge shape.
 
In this paper we develop theoretical models to describe the behaviour of two interacting magnetic-capillary micro-rafts. We identify three different regions of behaviour: the mean separations of the rafts, the collapse dynamics of the rafts, and the assembled configurations. These regions are each governed by different physics. Simple models are created for each region by considering the influence of magnetic, capillary and hydrodynamic interactions. Finally, we combine the simple models to describe the behaviour of a new series of experiments.
 
 This article is organised as follows. In Sec.~\ref{sec:experiment} we summarise the results of previous magnetic-capillary micro-rafts experiment \cite{Wang2017} and discuss the configuration of the new experimental configuration. Section~\ref{sec:physics} then considers the governing physics of the assembled raft configurations (Sec.~\ref{sec:Assemble}), the collapse dynamics (Sec.~\ref{sec:collapse}) and the mean separation (Sec.~\ref{sec:mean}). Finally in Sec.~\ref{sec:full} we join together the simple models and compare them to the results to a new series of  experiments.

\section{Experimental capillary disks} \label{sec:experiment}

\begin{figure*}[t]
\begin{center}
\includegraphics[width=\textwidth]{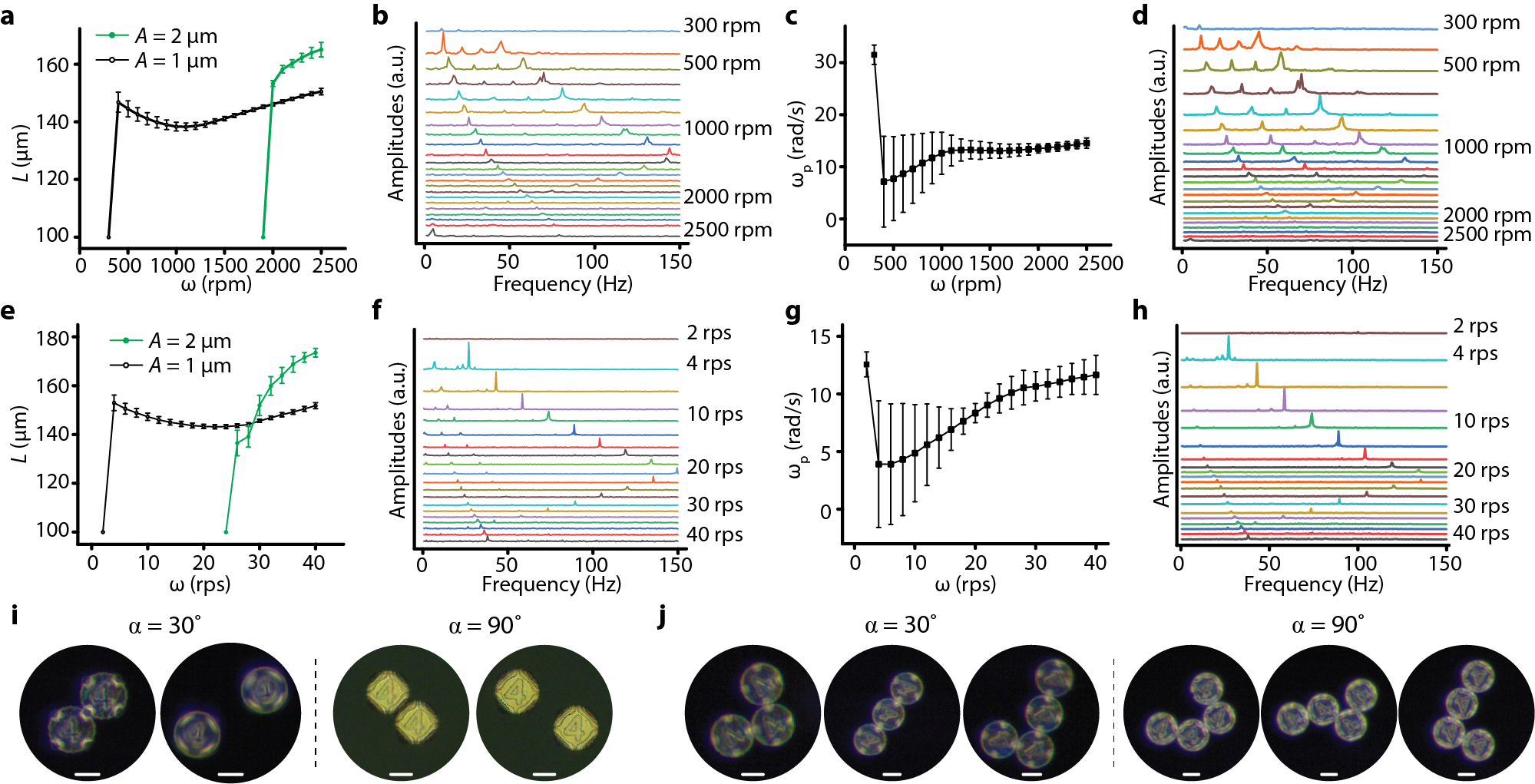}
\caption{Experimental results of the pairwise interactions of capillary micro-raft and configurations of static self-assembly for three to four micro-rafts. (a)-(d): Results of experiments previously reported in Ref.~\cite{Wang2017}. The structural parameter for these plots is $\alpha = 1/12$ and the frequency of the driving magnet is measured in revolutions per minute (rpm). (e)-(h): Results of new experiments conducted for this paper. The structural parameter for these plots is $\alpha = 1/4$ and the frequency of the driving magnet is measured in revolutions per second (rps). (a) and (e): Mean separation measured as a function of the driving frequency for both $A=1$ (black) and $2~\mu$m (green) rafts. (b) and (f): The Fourier transform of edge-to-edge distances observed for the $A=1~\mu$m rafts. (c) and (g): Mean orbital velocity of the two raft system as a function of driving frequency for the $A=1~\mu$m rafts. (d) and (e):  Fourier transform orbital velocity of the $A=1~\mu$m rafts. (i): images of assembled and orbiting pair of rafts for both $\alpha=1/12$ and $\alpha=1/4$. (j): Assembled configurations of three to four micro-rafts with $\alpha$ = 1/12 and $\alpha$ = 1/4 as observed in Ref.~\cite{Wang2017}.  Scale bars in panels (i) and (j) are 50~$\mu$m and images were adapted from Ref.~\cite{Wang2017} as permitted under the AAAS's License to Publish. } 
\label{fig:experimentdata}
\end{center}
\end{figure*}

In our experiments, a spinning permanent magnet (5mm cube NdFeB magnet N42) provides the rotating magnetic field used to rotate the two magnetic micro-rafts, of radius $R$=50~$\mu$m, floating at air-water interface. The structural parameters of a representative micro-raft and the current experimental set-up of which are shown in Fig.~\ref{fig:experiment}. The amplitude of the rafts, $A$, can be varied in the range $0 - 4~\mu$m, and the fraction of the edge that is occupied by one cosine, $\alpha$ (such that arc angle is $2 \pi \alpha$ in radians), can be varied between $0 - 1/4$. We show results with A = 1 and 2~$\mu$m and $\alpha$ = 1/12  (old experiments) and 1/4 (new experiments).

 In the previous experimental set-up \cite{Wang2017}, the magnet was driven by a laboratory stir plate (IKA Big Squid White), with the unit of rotation being revolutions per minutes (rpm). A high-speed camera (Basler acA800-510uc) was mounted on a stereo-microscope (Zeiss Discovery Z12), and equipped with a LED ring light illumination to record the motion of the micro-rafts. We have since improved our experimental setup to achieve finer control over the rotation and obtain better a image quality. Our new stepper-motor system (Lin Engineering 4118S stepper motor with R356 controller) offers better control of the rotation of the permanent magnet and uses revolutions per second (rps) as the unit. In addition, we have built an imaging system based on InfiniTubeTM Standard with in-line white light LED illumination to deliver better lighting conditions. This shortens the exposure time from 2 ms to $200 - 500~\mu$s and allows us to obtain sharper images while recording at high speed (300 fps). 

In the latest experiments, the quality of the magnetic thin film on each raft has also been increased. The cobalt thin films were sputtered onto the surface of the as-printed micro-rafts at the sputtering power of 100~W and the argon flow rate of 560~sccm. In the new preparation procedure, we prolonged the duration of the initial pumping such that the based vacuum is ~5$\times$10$^{-7}$~Torr, as compared with the previous vacuum pressure of ~1$\times$10$^{-6}$~Torr. This decrease in base vacuum pressure reduces the oxidizing species in the deposition chamber and increases the quality of the cobalt thin films.

Videos of the experimental pairwise interactions are analyzed with a custom Python code to first extract the positions and the orientations of micro-rafts over time. Based on the position information, we calculate the precession speed, the distances between micro-rafts, and the Fourier spectra of the distances and angular velocity over time. The results of these new experiments are displayed in Fig.~\ref{fig:experimentdata}.

A pair of rafts in this new set-up behaves similarly to the previously reported results but with a lower critical collapse frequency. At large magnetic driving frequency the rafts are observed to orbit around a central point with a well defined mean separation distance. As the driving frequency decreases, this mean separation decreases and the standard deviation on the mean increases. Eventually at a critical, non-zero, driving frequency the disks collapse and assemble into a static configurations (Fig.~\ref{fig:experimentdata}i). These static configurations depend on the edge profile assigned to each raft; rafts with $\alpha = 1/12$ only assemble into configurations where the peak of the edge profile of one raft connects the peak of the profile of another raft, hereafter denoted by max-max configuration, while rafts with $\alpha = 1/4$ assemble into configuration where either the peaks or troughs of the profiles are in contact, denoted max-max or min-min respectively (Fig.~\ref{fig:experimentdata}j). Furthermore rafts with $A=1~\mu$m also exhibit an increase in the mean separation just before collapse. The cause of this increase is unclear and, because it is not found for $A>1~\mu$m, will be left for later investigations. Importantly, when the rafts are flat they do not exhibit any orbiting behaviour.

\section{The physics of two interacting disks} \label{sec:physics}

The rich behaviour exhibited by the magnetic-capillary micro-rafts indicates that it is a complex physical processes. The orbital positions, the collapse dynamics, and the assembled configurations all show different non-linear dynamics. This suggests that they are each governed by different physical balances. Investigation into each of these three phases separately could therefore help explain the system as a whole. These investigation are best treated in reverse order, as the preceding behaviours require additional physics.

\subsection{Static self-assembly configurations} \label{sec:Assemble}

The simplest behaviour to address theoretically is the static configurations of the rafts. In this regime there is no dynamics and the behaviour is dominated by the surface capillary energy, as demonstrated by the dependence of the configuration on the edge profile of the disks.  No dependence on the the magnetic moments of the rafts has been observed in these configurations. The capillary energy captures the effective cost to distort the air-liquid surface from a perfectly flat interface due to surface tension. 

Many groups have developed static self-assembling structures which use capillary interactions \cite{Bowden1999, Lewandowski2010, Botto2012, ANIE201005078, Poty2014} and so the associated energy has been studied extensively \cite{Vella2005, Yao2013a, Kralchevsky2000}, particularly in the small surface deformation limit. In this limit, the governing equations for the surface deformation become linear. Hence the total capillary energy for any system can be found by adding together the energy of the `modes' that make up the structure. In the case of circular rafts the modes are often chosen to be the modes of a Fourier series. This is because for any given edge undulation profile, $H(\theta)$, it is always possible to write it as
\begin{equation}
H(\theta) = \sum_{n=0}^{\infty} A_{n} \sin(n \theta),
\end{equation}
where $A_{n}$ are the Fourier coefficients and $\theta$ is the angular parametrisation. The total energy for such a boundary is then simply the sum of the energy of each mode.

\begin{figure}
\begin{center}
\includegraphics[width=0.45\textwidth]{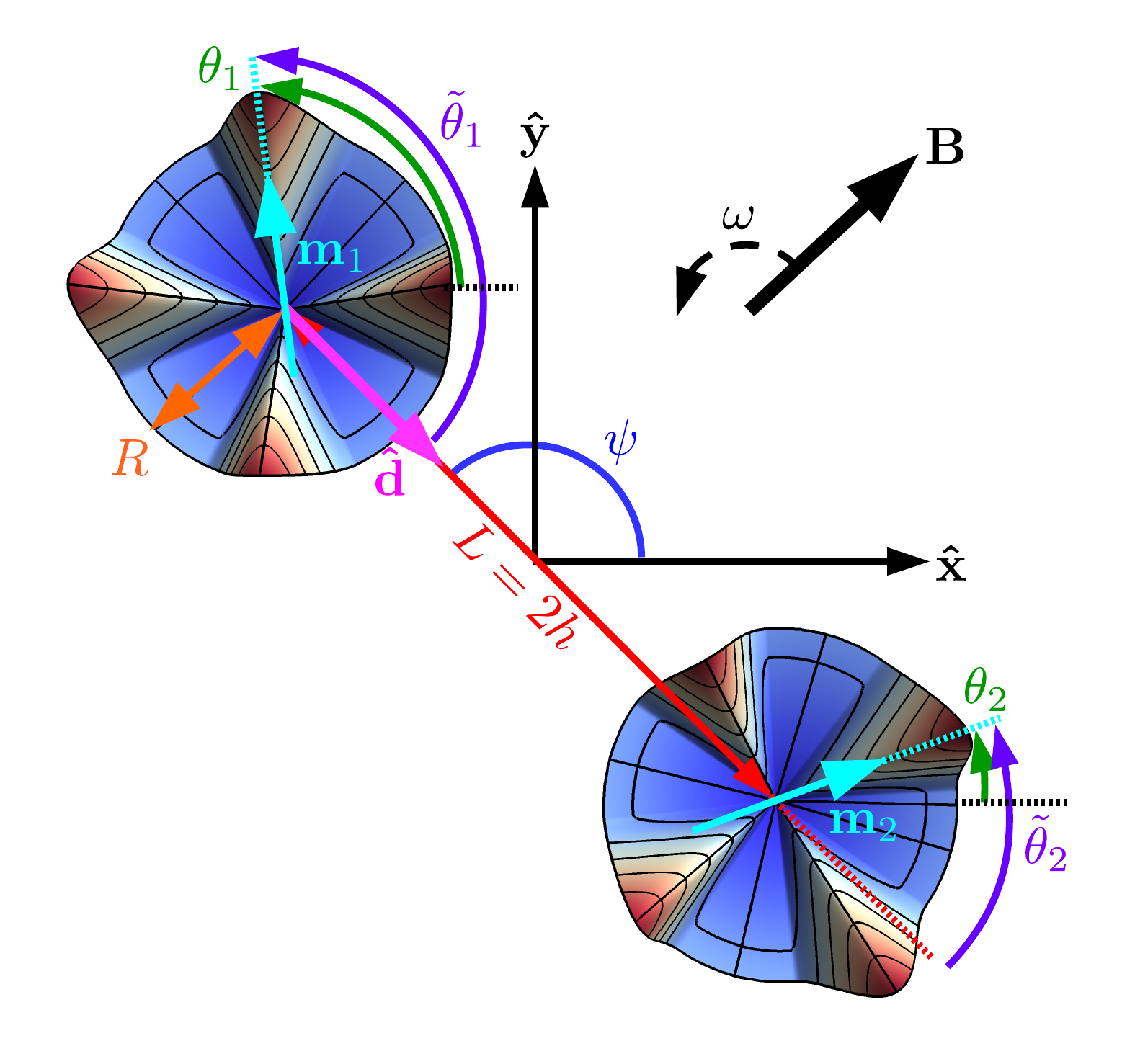}
\caption{ Diagram of two interacting rafts as viewed from above the surface. Here $B$ indicates the direction of the background field, $L=2h$ is the centre-to-centre separation of the two rafts, $\psi$ is the precession angle of the rafts relative to the laboratory frame, $\mathbf{m}_{i}$ is the magnetic moment of raft $i$, $R$ is the radius of the rafts, $\theta_{i}$ is the angular orientation of the magnetic moment on raft $i$ in the laboratory frame and $\tilde{\theta}_{i}$ is the relative angular orientation of raft $i$.}
\label{fig:diagram}
\end{center}
\end{figure}

For two circular disks with arbitrary sinusoidal edge profiles, the capillary energy can be calculated exactly \cite{Danov2005}. This was achieved by solving for the shape of the free surface in bipolar coordinates. The resultant surface energy, $E_{m_{1},m_{2}}$,  of two disks of radius $R$, separated by a distance $L$, is
\begin{equation}
\frac{E_{m_{1},m_{2}}}{\pi \sigma} = H_{1}^{2} S_{m_{1}} + H_{2}^{2} S_{m_{2}} - H_{1} H_{2} G_{m_{1},m_{2}} \cos(m_{1} \tilde{\theta}_{1} + m_{2} \tilde{\theta}_{1}),
\end{equation}
where $m_{i}$ is the sinusoidal mode of raft $i$, $H_{i}$ is the amplitude of the sinusoid on raft $i$, $\sigma$ is the surface tension of the air-fluid interface, $\tilde{\theta}_{i}$ is the angle between the vector connecting the rafts and a sinusoidal maxima of disk $i$ and where $i=1,2$ indicates the disk number (see notation in Fig.~\ref{fig:diagram}).
In the above equation $S_{n}$ and $G_{n,m}$ are summation factors related to the geometry and are given by
\begin{eqnarray}
S_{n} &=& \sum_{k=1}^{\infty} \frac{k}{2} \coth\left[2 k \mbox{ acosh}\left(\frac{L}{2 R}\right)\right] \Xi^{2}\left[k,n, \mbox{acosh}\left(\frac{L}{2 R}\right)\right] ,\notag \\
\\
G_{n,m} &=& \sum_{k=1}^{\infty} \frac{k \Xi\left[k,n, \mbox{acosh}\left(\frac{L}{2 R}\right)\right] \Xi\left[k,m, \mbox{acosh}\left(\frac{L}{2 R}\right)\right]}{\sinh\left[2 k \mbox{ acosh}\left(\frac{L}{2 R}\right)\right]},\\
\Xi(n,m,\nu) &=& m \sum_{k=0}^{\min(m,n)}\frac{(-1)^{m-k}(m+n-k-1)!}{(m-k)!(n-k)!k!} e^{-(m+n-2k) \nu},
\end{eqnarray}
where $\mbox{acosh}(x)$ is the inverse of the hyperbolic cosine function \cite{Danov2005}. These solutions are the natural mode decomposition for two interacting circular rafts and the total energy, $E$, of any given pair of undulations simply the sum of all the different energy modes,
\begin{equation}
E = \sum_{m_{1},m_{2}} E_{m_{1},m_{2}}.
\end{equation}

The experimental mirco-raft system use cosine shaped undulations separated by flat regions. The Fourier series representation of this edge is  written as
\begin{equation}
H(\theta) = \sum_{n=1}^{\infty} \frac{2 A}{\pi n (1 -16 n^{2} \alpha^{2})} \sin\left(4 n \pi \alpha \right) \cos(4 n \theta),
\end{equation}
where $H(\theta)$ is the  height of the raft edge at the angle $\theta$, $A$ is the amplitude of the raft edge and $\alpha$ is the fraction of the edge that is occupied by one cosine (Fig.~\ref{fig:experiment}a). These Fourier coefficients decay with $n \alpha$ and so require more terms when $\alpha$ is small. From this Fourier representation, the total capillary energy  $E$ between two rafts is
\begin{multline}
\frac{\pi E}{4 A^{2} \sigma} = 2 \sum_{n=1}^{\infty} \left(\frac{\sin\left(4 n \pi \alpha \right)}{ n (1 -16 n^{2} \alpha^{2})} \right)^{2} S_{4 n} \\
 - \sum_{n,m=1}^{\infty} \frac{ \sin\left(4 n \pi \alpha \right) \sin\left(4 m \pi \alpha \right) G_{4n,4m}}{n m (1 -16 n^{2} \alpha^{2}) (1 -16 m^{2} \alpha^{2})}  \cos[4(n \tilde{\theta}_{1}+ m \tilde{\theta}_{2})], \label{E}
\end{multline}
where   $\tilde{\theta}_{1}$ (resp.~$\tilde{\theta}_{2}$) is the angle between the vector connecting the two rafts and the vector pointing to a maxima of raft one (resp.~two). Without loss of generality we chose this maxima to coincide with the direction of the magnetic moment on the raft (Fig.~\ref{fig:diagram}). 

Though the above energy is exact it can be hard to work with. Hence it is typical to consider the far-field behaviour of this function, i.e.~when $L \gg R$. In this limit, the energy reduces to the asymptotic value $E_{far}$ given by\cite{Danov2005}
\begin{multline}
E_{far} \approx - \frac{8  \sigma A^{2}}{\pi}\sum_{n,m=1}^{\infty} \left\{ \frac{  (4 n+4 m -1)!}{(4 n-1)!(4 m-1)!} \right. \\
\left.\times \frac{\sin\left(4 m \pi \alpha \right) \sin\left(4 n \pi \alpha \right)}{ n m (1 -16 n^{2} \alpha^{2}) (1 -16 m^{2} \alpha^{2})}  \frac{ R^{4(m+n)} \cos\left[4 (n \tilde{\theta}_{1} + m \tilde{\theta}_{2}) \right] }{ L^{4(m+n)}} \right\}.\label{Efar}
\end{multline}
 This far-field energy underestimates the value of the actual surface energy as the raft separation decreases but it retains the correct dependence on the rafts orientation \cite{Danov2005} and is much easier to evaluate. This angular dependence determines the assembled configurations.

\begin{figure}[t]
\begin{center}
\includegraphics[width=0.4\textwidth]{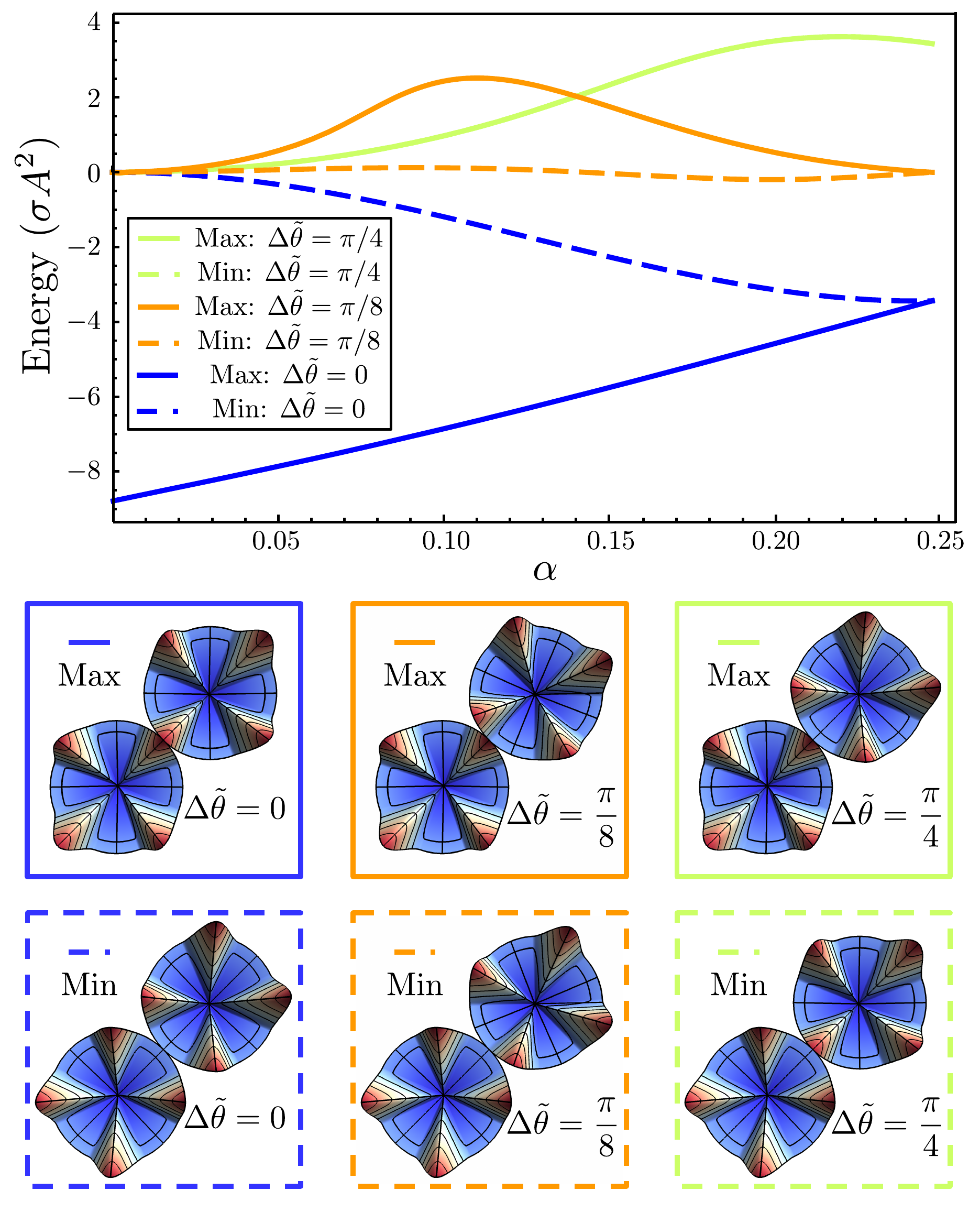}
\caption{The far-field surface energy of two capillary disks in contact for different relative configurations and $\alpha$. The solid lines show the energy when the maximum of the first disk is held in contact with the other disk while the dashed lines show the energy when the minimum is held in contact. Illustrations bellow show the two-disk configuration for each line at $\alpha=1/7$. The two lines for $\Delta\tilde{\theta} = \pi/4$ (light green) correspond to the same configuration and so are identical. These far-field results capture the relative energy of the configurations but the full surface energy, Eq.~\eqref{E}, is required for a quantitative prediction of the contact energy.}
\label{fig:energy}
\end{center}
\end{figure}

The far-field energy with $L=2R$ is plotted in Fig.~\ref{fig:energy}  for $\tilde{\theta}_{1}=0$ (max configuration) and  $\tilde{\theta}_{1} = \pi/4$ (min configuration) with $\tilde{\theta}_{2} = \tilde{\theta}_{1} +\Delta \tilde{\theta}$. As expected these energy curves coincide when $\Delta \tilde{\theta}=\pi/4$, corresponding to a maximum of one disk touching the minimum of the other. Furthermore the energy of this state is maximised near $\alpha=1/4$ and decreases as the width of the undulation decreases. This decrease reflects the amount of overlap between the low and high parts of the edge profile as $\alpha$ decreases, with the slight increase at large $\alpha$ probably arising from the additional multipoles needed to represent the edge profile.

As $\Delta \tilde{\theta}$ decreases, the energy of the maximum (solid) and minimum (dashed) contacts begin to differ. This difference can be significant, with the energy of the maximum configuration being larger than the $\Delta \tilde{\theta} = \pi/4$ energy for $\Delta \tilde{\theta}=\pi/8$ and $\alpha \sim 0.125$. The energy of the minimum configuration is effectively 0 over this region. This energy cost can again be attributed to the overlap of the edge profiles. For $\alpha = 0.125$, exactly half the edge of each raft is flat. Therefore when $\Delta \tilde{\theta}=\pi/8$, the maximum of the second raft sits at the start of a flat region. This creates a large amount of unfavourable overlap, which decreasing as $\alpha$ increases and decreases.

Finally, when $\Delta \tilde{\theta} =0$, the energy of a maximum-maximum (max-max) contact (blue solid) and a minimum-minimum (min-min) contact (blue dashed) is considered. The energy for the max-max configuration is always lower than its min-min counterpart except when in the octopole configuration, $\alpha=1/4$. This difference in energy therefore explains why $\alpha=1/12$ disks were only found in the max-max configuration while $\alpha=1/4$ disks could be found in both max-max and min-min contact. The energy difference between these two configurations arises from the interaction of the Fourier modes in $E_{far}$. In the max-max configuration all the Fourier modes reduce the energy of the state. However in the min-min state only the modes with even $m+n$ reduce the energy, causing it to have a higher overall energy. As $\alpha$ decreases, the energy gap increases due to the increase in the number of modes needed to represent the overall shape. Finally, when $\alpha=0$, the undulation becomes an infinitely thin peak, and so the min-min state becomes the energy of two flat disks while the max-max state becomes that of aligned delta-like functions.

\subsection{Collapse dynamics} \label{sec:collapse}

For our microscopic rafts, the ratio of inertial to viscous forces, i.e.~the Reynolds number, is $Re= \rho R^{2} \omega/ \mu \sim 0.01-0.1$, where $\rho$ is the density of the fluid, $\mu$ is the dynamic viscosity and $\omega$ is the driving frequency.  
 Hence the disks are approximately force and torque free and their dynamics is dictated by the balance of viscous drag with the capillary and magnetic interactions. For separated rafts, the torque balance is dominated by the interaction between the raft's magnetic moment and the rotating permanent magnet, which scales as $m B \sim 10^{-12}$~Nm, since the magnetic dipole-dipole torque between the rafts scales as $ \mu_{0} m^{2} / R^{3} \sim 10^{-13}$~Nm and  the capillary torque as $ \sigma A^{2} \sim 10^{-13}$~Nm \cite{Wang2017}, where $\mu_{0}$ us the permeability of free space. As a result, the torque balance on each raft in the direction perpendicular to the interface is approximately given by
\begin{equation}
0 = L_{M} - L_{v}  = (\mathbf{m}\times \mathbf{B})\cdot\mathbf{\hat{z}}  - L_{v}\approx m B_{0} \sin[\omega t-\theta_{i}(t)] - L_{v}, \label{torque}
\end{equation}
where $L_{v}$ is the viscous torque, $ L_{M}$ is the torque from the background magnetic field, $\mathbf{m} = m \{\cos(\theta_{i}), \sin(\theta_{i}),0\}$ is the magnetic dipole moment of raft $i$ ($i=1,2$), $\mathbf{B} = B(r) \{\cos( \omega t), \sin( \omega t),0\}$ is the background magnetic field, $\mathbf{\hat{z}}$ is the normal to the air water interface, $B(r)$ is the strength of the background field as a function of the distance from the centre of the magnet, $r$, $\omega$ is the rotation frequency of the background field, $t$ is the time, and $\theta_{i}$ is the orientation of the magnetic moment of raft $i$ relative to the laboratory frame (see notation in Fig.~\ref{fig:diagram}). In Ref.~\cite{Wang2017} the magnetic field was weakly quadratic, $B(r)\approx B_{0} + r^{2} B_{2}$, where $B_{0}$ and $B_{2}$ are constants. Experiments found $B_{0}=20\times10^{-3}$~T and $B_{2}r^{2} \sim  B_{2} R^{2} = 5.4 \times 10^{-6}$~T, where $B_{2}= 2.15 \times 10^{3}$~Tm$^{-2}$. The contributions from the $B_{2}$ terms are therefore negligible for the raft rotations.

In contrast with the torque balance, all  forces acting on the rafts are expected to be of similar magnitude. However, inspecting the Fourier transforms of the experimental data, Figs.~\ref{fig:experimentdata}f and h, the largest peak is at eight times the driving frequency. This eight times peak is due to fourfold symmetry in edge undulations. Hence the capillary force must play a significant role to the dynamics and so the force balance on each raft is 
\begin{equation}
\mathbf{0} = \mathbf{F}_{c,far} - \mathbf{F}_{v} = - \partial_{L} E_{far} \mathbf{\hat{d}} - \mathbf{F}_{v} = -2240 \pi \sigma A^{2} R^{8}  \frac{ \cos\left(8 \tilde{\theta}\right) }{ L^{9}}  \mathbf{\hat{d}} - \mathbf{F}_{v}, \label{force}
\end{equation}
where  $\mathbf{F}_{v}$ is the viscous drag force, $\mathbf{F}_{c,far}$ is the far-field component of the capillary force, and $ \mathbf{\hat{d}} = \{ \cos[\psi(t)],\sin[\psi(t)],0\}$ is a unit vector pointing from the centre of raft one to raft two (see Fig.~\ref{fig:diagram}). In the above we have already assumed that $\alpha=1/4$ and $\tilde{\theta}_{1} =\tilde{\theta}_{2} =\tilde{\theta}$.

With the force and torque balance known, the movement of the rafts depends   then critically on the  viscous drag terms. At low Reynolds number, the drag forces and torques are linearly related to the linear and angular velocity of each raft and only depend on the instantaneous shape and relative distance of the system. For simplicity we assume that the disks remain flat on the air-water interface throughout the motion (so the motion is two dimensional), and that they can be described by oblate ellipsoids. Under these conditions the drag force and torque (perpendicular to the surface)  experienced by an isolated raft are given by
\begin{eqnarray}
2\mathbf{F}_{v} &=&  \frac{ 16 \pi \mu Re^{3}}{(2 e^{2} +1) C - e \sqrt{1-e^{2}}} \mathbf{U} \label{forcev} , \\
2 L_{v} &=& \frac{16}{3} \frac{ \pi \mu R^{3} e^{3}}{ C - e \sqrt{1-e^{2}}} \Omega  \label{torquev},
\end{eqnarray}
where $e = \sqrt{1-z^{2}/R^{2}}$ is the eccentricity of the raft, $C = \mbox{arccot}(\sqrt{1-e^{2}}/e)$, $z= 2.5~\mu$m is half the thickness of the rafts, $\mathbf{U}$ is the velocity in the plane, and $\Omega$ is the raft angular velocity along the normal to the surface  \cite{Kim2004}. The factor of two on the left hand side of the above equations reflect that only half the ellipsoids are submerged in the fluid, and we ignore any drag coming from the air motion above it.

 In general the correction to the forces and torques for the hydrodynamics interaction of the oblate ellipsoids is very complex, even when the bodies are far from each other \cite{Yoon1990}. We may approximate these corrections by the those for two  far-separated spheres, which are simple and readily available from the literature~\cite{Kim2004}. These spherical leading-order corrections have the same dependence on the distance between the bodies but with a slightly different pre-factors and so are a suitable approximation for the purpose of understanding the governing physics in the experiments. Hence in the far-field limit the approximate velocity-drag relationships become
\begin{eqnarray}
\mathbf{U} &=& \frac{(2 e^{2} +1) C - e \sqrt{1-e^{2}}}{ 8 \pi \mu Re^{3}} \left(1-\frac{3 R}{2 L}\right)\mathbf{F}_{v}\notag \\
&& - \frac{  C - e \sqrt{1-e^{2}}}{\pi \mu R^{2} e^{3}}\frac{3 R^{2}}{4L^{2}} L_{v} \mathbf{\hat{d}}\times\mathbf{\hat{z}},   \label{vel} \\
\Omega &=& - \frac{  C - e \sqrt{1-e^{2}}}{\pi \mu R^{2} e^{3}}\frac{3 R^{2}}{4L^{2}} (\mathbf{\hat{d}}\times\mathbf{F}_{v})\cdot\mathbf{\hat{z}} \notag \\
&&+ \frac{3}{8} \frac{ C - e \sqrt{1-e^{2}} }{\pi \mu R^{3} e^{3}} L_{v}, \label{rot}
\end{eqnarray}
where $L$ is the separation between the rafts (Fig.~\ref{fig:diagram}). The coupling term between the torque and velocity accounts for the drift experienced by one raft, $U \propto R^{3} \Omega/ L^{2}$, when placed in the flow field of the second raft rotating with angular velocity $\Omega$. The equivalent couple, between force and angular velocity, is then required as per the symmetries of viscous hydrodynamics \cite{Kim2004}. Similarly the correction to the velocity-force relation, given  by a factor $1-3R/2L$, accounts for the additional drag on the disk due to the flow created by the second disk with a force $-\mathbf{F}_{v}$. The combination of   Eqs.~\eqref{vel} and \eqref{rot} with the force and torque balances from Eqs.~\eqref{torque} and \eqref{force} determines the linear and angular velocity of each raft.

A suitable parametrisation of the raft configuration is needed to relate these velocities to their physical motion. For the system of identical rafts we  consider, this configuration is uniquely defined by three parameters:  the orientation of the raft relative to the lab frame, $\theta=\theta_{1}=\theta_{2}$,  the distance of the raft from the origin, $h$ (so that $L=2h$), and  the angle $\psi$ between the axis connecting the rafts and the laboratory frame (i.e.~the precession angle). These parameters are illustrated in Fig.~\ref{fig:diagram}. In this parametrisation the vector from raft one to raft two takes the form $L \mathbf{\hat{d}} = 2 h \{ \cos[\psi(t)],\sin[\psi(t)],0\}$ with the identity $\tilde{\theta} = \theta(t)-\psi(t) +\pi$ resulting from geometry. As the director between the rafts, $\mathbf{\hat{d}}$, is present in both the viscous drag and the force balance, the motion of the rafts  is naturally separated into its motion along $\mathbf{\hat{d}}$ and that perpendicular to it. Hence the equations describing the evolution of the each raft are given by
\begin{eqnarray}
\frac{d \theta}{d t} =& \Omega &= \beta \sin[t-\theta(t)], \label{pos3} \\
\frac{d h}{d t} =& \mathbf{U}\cdot\mathbf{\hat{d}} &=  -\frac{35 \gamma}{6}\left( 1- \frac{3}{4 h }\right)  \frac{ \cos\left[8 \left(\theta(t)-\psi(t) +\pi\right)\right] }{ h^{9}}, \notag \\
 \label{pos1} \\
h \frac{d \psi}{ d t} =& \mathbf{U}\cdot\mathbf{\hat{n}} &=  \frac{\beta}{ 4 h^{2} }  \sin[ t - \theta(t)]  \label{pos2},
\end{eqnarray}
where $\mathbf{\hat{n}} = \{-\sin[\psi(t)],\cos[\psi(t)],0\}$, $\beta = 3 m B_{0} ( C - e \sqrt{1-e^{2}}) / 8 \pi \mu \omega R^{3} e^{3}$ is the ratio of the magnetic and viscous torques, and $\gamma = 3 \sigma A^{2} \left[ (2 e^{2} +1) C - e \sqrt{1-e^{2}}\right]/ 32 \mu \omega R^{3} e^{3} $ is an inverse capillary number. In the above we have scaled length by $R$ and time by $1/\omega$.

The equation for the rotation of the raft, Eq.~\eqref{pos3}, has an exact solution when $\beta \geq 1$, namely
\begin{equation}
\theta(t) = t - \arcsin\left( \frac{1}{\beta}\right). \label{theta}
\end{equation}
 Our model rafts therefore rotate linearly with the background magnetic field.  This is consistent with the experimental observation of negligible rotation fluctuations around a mean velocity \cite{Wang2017}. The equations for the separation of the rafts, Eq.~\eqref{pos1}, and the procession angle, Eq.~\eqref{pos2}, however do not have a closed form solution and so must be evaluated numerically. 

\begin{figure}
\begin{center}
\includegraphics[width=0.45\textwidth]{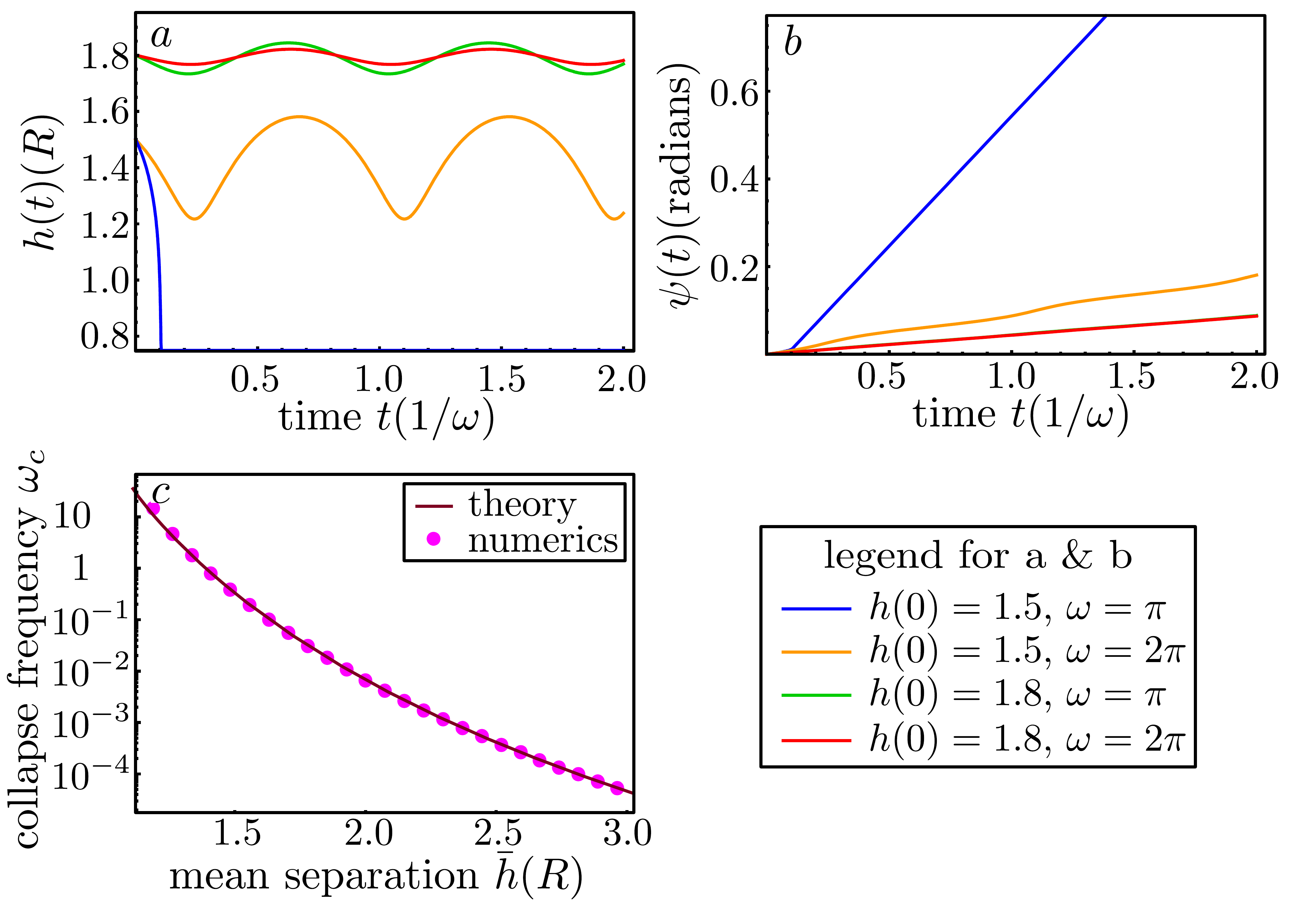}
\caption{Numerical solutions  of Eqs.~\eqref{pos1}, and \eqref{pos2}. 
(a): Half the centre-to-centre separation, $h$, as a function of time. 
(b): Procession variable, $\psi$, as a function of time. 
(c): Relationship between mean position and the critical driving frequency as determined numerically and approximately. In these figures we used the parameters $\beta=318.6/\omega$ and $\gamma=74/\omega$.}
\label{fig:dynamic2}
\end{center}
\end{figure}

We evaluated the separation and procession angle equations, Eqs.~ \eqref{pos1}, and \eqref{pos2}, using Mathematica's NDSolve function \cite{Mathematica}. The solutions for various initial heights and background driving frequencies, $\omega$, are plotted in Fig.~\ref{fig:dynamic2}a and b.
When $\omega$ is large, the solution forms a stable oscillation around some mean value $\bar{h}$. As $\omega$ decreases these oscillations get larger until eventually, at a critical driving frequency, $\omega_{c}$, the rafts touch and the system collapses. In our model, the rafts collapse to $h=3/4$ due to the leading-order far-field hydrodynamic corrections. In experiments the increasing error around the mean position \cite{Wang2017} and the Fourier spectrum of the position (Fig.~\ref{fig:experimentdata}b and f) also indicate that the oscillations around the mean grow as the driving frequency decreases. Notably, in the Fourier spectrum the increasing peaks sit near 2x and 8x the driving frequency. This suggests that the collapse dynamics involves both capillary interactions (the 8x peak) and magnetic dipole-dipole interactions (the 2x peak). The smaller peaks in the spectrum could relate to other magnetic multipole interactions but appear to be negligible.

 In our simple model, the critical collapse frequency, $\omega_{c}$, and the mean separation, $\bar{h}$, depend on the initial separation of the rafts, $h(0)$ (Fig.~\ref{fig:dynamic2}a). This is in contrast with the experiment, in which $\omega_{c}$ and $\bar{h}$ is independent of the initial separation. This is because the mean separation is governed by different physics, as we  discuss in the next section. The current collapse model, however, can be used to create a relationship between $\bar{h}$ and $\omega_{c}$, thereby providing an approximate collapse frequency for any observed mean separation. To construct this relationship we assume the collapse occurs when $h(t)=1$, the mean position can be approximated by $\bar{h} \approx (h(0) + 1)/2$ and that $\psi(t)$ is approximately linear with time and so takes the form $\psi(t)  \approx t/4 \bar{h}^{3}$. These approximations are empirically deduced from  the numerical results in Fig.~\ref{fig:dynamic2}a and b. With these assumptions, the equation for the separation of the rafts, Eq.~\eqref{pos1}, becomes
\begin{equation}
\frac{d h}{d t} = -\frac{35  \gamma}{6}\left( 1- \frac{3}{4 h }\right)  \frac{ \cos\left[8 \left(1-\frac{1}{4 \bar{h}^{3}} \right) t - 8 \arcsin\left(\frac{1}{\beta}\right)\right] }{ h^{9}}\cdot
\end{equation}
This is a separable differential equation with   solution
\begin{multline}
-\frac{48(4 \bar{h}^{3}-1)\left(f[h]-f[h(0)]\right)}{35 \bar{h}^{3} \gamma} =\\
  \sin\left[2 \left(4 + \frac{1}{\bar{h}^{3}}\right)t + 8 \arcsin\left(\frac{1}{\beta}\right)\right] - \sin\left[ 8 \arcsin\left(\frac{1}{\beta}\right)\right] , \label{simplesol}
\end{multline}
where 
\begin{multline}
f[h] = \frac{19683 h}{1048576}  + \frac{6561 h^{2}}{524288}  + \frac{729 h^3}{65536} + \frac{729 h^4}{65536} + \frac{243 h^5}{20480} + \frac{27 h^6}{2048} \\
+ \frac{27 h^7}{1792} + \frac{9 h^8}{512} + \frac{h^9}{48} + \frac{h^{10}}{40} + \frac{59049 \ln(3 - 4 h)}{4194304}.
\end{multline}
The right-hand side of Eq.~\eqref{simplesol} is the sum of two sinusoidal functions. The second these these sinusoids however is independent of time and is very small for typical values of $\beta$ ($\sim 318.6/\omega$). Hence solutions to the above equation only exist if the left hand side is between $-1$ and $1$. 
Recalling that collapse occurs when $h=1$, the criteria for collapse becomes
\begin{equation}
\left| \frac{48(4 \bar{h}^{3}-1)\left(f[1]-f[2 \bar{h} -1]\right)}{35 \bar{h}^{3} \gamma} \right| = 1,
\end{equation}
which, because $\gamma$ depends inversely on $\omega$, provides a relationship between the mean position, $\bar{h}$, and collapse frequency, $\omega_{c}$.

This theoretical prediction is plotted in Fig.~\ref{fig:dynamic2}c alongside the results from the numerical implementation of Eqs.~\eqref{pos1} and \eqref{pos2}. The theoretical relationship shows excellent agreement with the computational results. In addition to providing an approximate value of $\omega_{c}$ for any observed $\bar{h}$, this relationship also shows  that $\omega_{c}$ increases quickly as $\bar{h}$ decreases. This is because a greater mean separation requires a lower driving frequency to make a comparable sized oscillation.

\subsection{Mean orbital separation} \label{sec:mean}

\begin{figure}
\begin{center}
\includegraphics[width=0.45\textwidth]{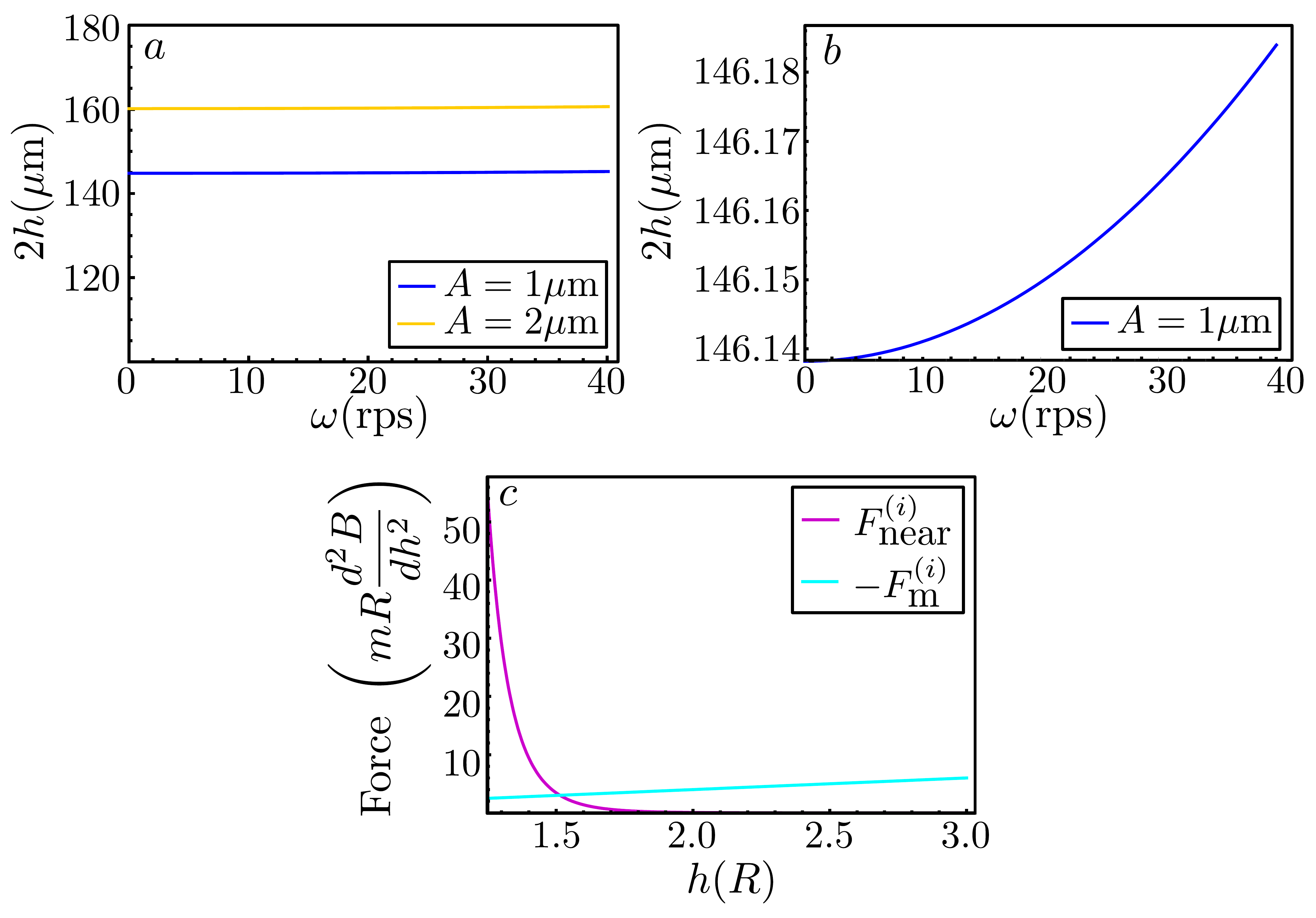}
\caption{Mean separation between the two micro-rafts as predicted by Eq.~\eqref{mean} as a function of $\omega$ for the current experimental regime. (a): Results for both $A=1~\mu$m and $A=2~\mu$m; (b) A close inspection of the $A=1~\mu$m separation showing a weak increase with $\omega$. (c) Strengths of the capillary near-field repulsion and magnetic attraction  as a function of $h$.  In all figures $m=2 \times 10^{-10}$~Am$^{2}$, $\sigma=74 \times 10^{-3}$~Nm$^{-1}$, $\mu= 10^{-3}$~Pas,  $B_{2}=2.15 \times 10^{3}$~Tm$^{-2}$, $B_{0}=20 \times 10^{-3}$~T and $R=5 \times 10^{-5}$~m.}
\label{fig:mean}
\end{center}
\end{figure}

The far-field capillary interaction is, on average, neither attractive nor repulsive and so cannot predict the mean separation between the disks. Experiments show however that capillary interactions are critical for the orbital configurations. This apparent paradox can be resolved by returning to the full capillary energy, Eq.~\eqref{E}. In this energy there are two distinct terms: one term that is independent of the configuration but is associated with the presence of a near by disk and one that depends on the relative orientation of the disks through a cosine. In the far-field approximation, the former of these terms is asymptotically smaller than the latter and so is often ignored \cite{Danov2005}. However, when the disks are rotated, the far-field contribution averages to zero over one period and therefore the near-field component (which does not average to zero)  becomes the leading contribution to the force. In our raft system these near-field terms create an orientation-independent repulsive force between the disks of the form
\begin{equation}
{\mathbf{F}}_{\rm near} = - \frac{8 \sigma A^{2}}{\pi} \sum_{n=1}^{\infty} \left(\frac{\sin\left(4 n \pi \alpha \right)}{ n (1 -16 n^{2} \alpha^{2})} \right)^{2} \left. \frac{d S_{4n}}{d L}\right|_{L=2h} \mathbf{\hat{d}} \label{Fnear},
\end{equation}
where $\mathbf{\hat{d}}$ is the unit vector pointing from disk one to the disk two (see Fig.~\ref{fig:diagram}). In order to create stable orbits this repulsive term must balance an orientation-independent attractive force. For the micro-rafts system, this attraction comes from the quadratic part of the magnetic field exerted on each raft by the rotating permanent magnet. Since the strength of the field is approximately $B_{0} + h^{2} B_{2}$, this magnetic force has the form
\begin{equation}
\mathbf{F}_{\rm m} = \left(\mathbf{m} \cdot \nabla \right) \mathbf{B}  =2 m h B_{2} \cos[\theta(t)-t] \mathbf{\hat{d}} = 2  m h B_{2} \cos\left[\arcsin\left(\frac{1}{\beta}\right)\right] \mathbf{\hat{d}}, \label{Fmag}
\end{equation}
where $B_{2}$ is   constant~\cite{Wang2017} and we have assumed that the centre of the disks lies directly above the driving magnet. The mean position of the rafts is then determined by the balance
\begin{equation}
\mathbf{F}_{\rm m}+\mathbf{F}_{\rm near} =\mathbf{0}. \label{mean}
\end{equation}
Due to the complexity of the term $S_{4n}$, 
this equation must be solved numerically. Typical curves for the near-field capillary force, Eq.~\eqref{Fnear}, and the magnetic attraction, Eq.~\eqref{Fmag}, are shown in Fig.~\ref{fig:mean}c. As expected Eq.~\eqref{Fmag} shows a linear increase while Eq.~\eqref{Fnear} decreases monotonically allowing for a single mean separation. This separation is plotted in Fig.~\ref{fig:mean}a and b for varying driving frequencies and edge amplitudes. The values  predicted by the model are close to the mean positions seen in the experiments (Fig.~\ref{fig:experimentdata}e). Interestingly the mean separation weakly increases with the driving frequency (Fig.~\ref{fig:mean}b). This increase with $\omega$ is caused by the phase lag between the rafts magnetic moment and the rotating magnet due to viscous drag (i.e.~the $\beta$ term in the magnetic force) and can become significant for small values of $\beta \omega$.

\section{Complete dynamic model} \label{sec:full}

\begin{figure*}
\begin{center}
\includegraphics[width=\textwidth]{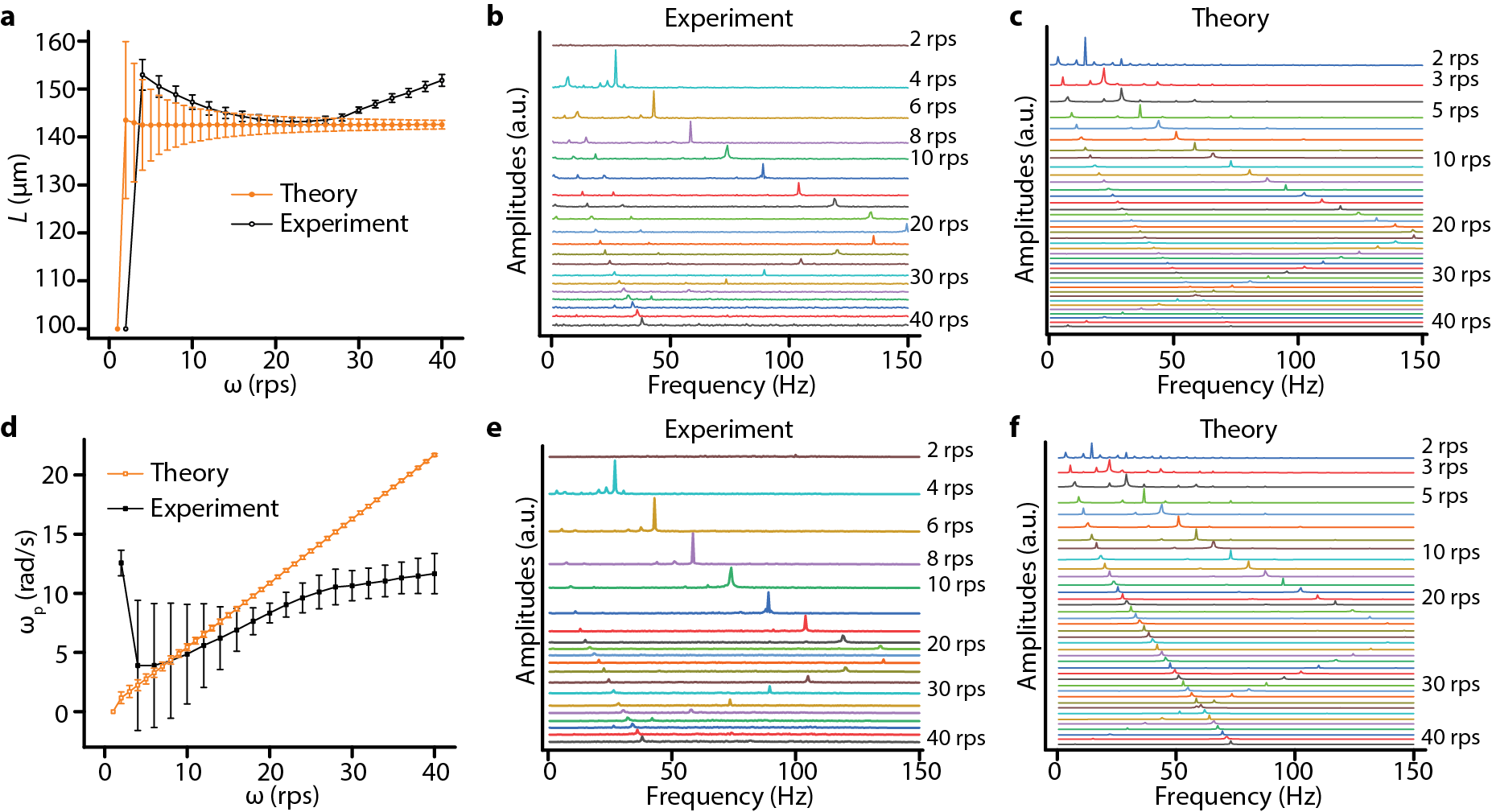}
\caption{Comparison between theoretical calculation and experimental results. (a) Mean separation between the disks, $L=2h$, as a function of driving frequency.  (b) and (c): Fourier spectrum of the rafts separation for the experiment and model respectively. (d) Mean precession frequency of the rafts, $\omega_{p} = \left\langle  {d \psi}/{d t} \right\rangle$, as a function of driving frequency, $\omega$.  (e) and (f):  Fourier spectrum of the precession frequency for the experiment and model respectively.}
\label{fig:full}
\end{center}
\end{figure*}

In the  previous sections we explored the governing physics of the assembled configurations, the collapse dynamics, and the mean separation separately and created a model to describe each behaviour in sequence. Here we combine these models and compare the results it to a new series of experiments with $A=1~\mu$m. We remind the reader that this experiment used rafts with a pure octopole undulations ($\alpha=1/4$) and were driven by a magnet on a stepper motor. 

This combined model is built identically to the collapse model in Sec.~\ref{sec:collapse}. However in the force balance equations, Eq.~\eqref{force}, the far-field capillary force is replaced with the complete capillary force, $\mathbf{F}_{c} = - \left(\partial_{L}E \right)\mathbf{\hat{d}}$ where $E$ is given by Eq~\eqref{E},
and the magnetic confinement force, Eq.~\eqref{Fmag}, is added. Furthermore the magnetic dipole-dipole force in the $\mathbf{\hat{d}}$ direction, given by
\begin{equation}
F_{dipole} = \frac{ 3 \mu_{0} m^{2}}{64 \pi R^{4}}\frac{1 +3\cos\left[2 \left(\theta -\psi \right)\right] }{h^{4}}\mathbf{\hat{d}},
\end{equation}
is included in order to capture the collapse frequency. This new force balance leaves the equations for the procession angle, Eq.~\eqref{pos2}, and the raft orientation, Eq.~\eqref{pos3}, unchanged but modifies the equation for the raft separation to
\begin{multline}
\frac{d h}{d t} = -\left( \frac{4}{3}- \frac{1}{ h }\right) \left[ \gamma \left\{2 \frac{d S_{4}}{d L} - \frac{d G_{4,4}}{d L}\cos\left[8 \left(\theta -\psi \right) \right] \right\} \right.\\
\left.+ \kappa h \cos\left(t-\theta\right) + \eta \frac{1 +3\cos\left[2 \left(\theta -\psi \right)\right] }{h^{4}} \right] ,
\end{multline}
where 
\begin{equation}
\kappa=3 m B_{2} \frac{(2 e^{2} +1) C - e \sqrt{1-e^{2}}}{16 \pi \mu \omega R e^{3} }
\end{equation}
 is the ratio of the magnetic attraction to the viscous drag while
 \begin{equation}
\eta =  9\mu_{0} m^{2} \frac{ (2 e^{2} +1) C - e \sqrt{1-e^{2}}}{ 2048 \pi^{2} \mu \omega R^{6} e^{3}}
\end{equation}
  is the ratio of the magnetic dipole-dipole force to the viscous drag. Note that we have dropped the explicit time dependence of the functions for notation simplicity. 

  This system of dynamic equations is then determined by four dimensionless parameters: $\beta$, $\gamma$, $\kappa$ and $\eta$. Although we used an improved deposition condition for magnetic cobalt thin film in the new series of experiments (Sec.~\ref{sec:experiment}), without additional knowledge we will assume that  the magnetic  moment has a similar value to that of Ref.~\cite{Wang2017}. 
The new  experimental results with $A=1~\mu$m are plotted in Fig.~\eqref{fig:full} alongside the results of our theoretical model. In the theoretical model we have $m= 2 \times 10^{-10}$~Am$^{2}$, $B_{0} = 20 \times 10^{-3}$~T, and $B_{2} = 2.15 \times 10^{3}$~Tm$^{-2}$ (as estimated in Ref.~\cite{Wang2017}), which corresponds to scaled parameters of $\gamma=186.7/\omega$, $\beta=3612/\omega$, $\kappa=1.727/\omega$ and $\eta=4.819/\omega$ where $\omega$ is in seconds. 
The theoretical results were obtained using Mathematica's NDSolve \cite{Mathematica}. The results were then sampled at 300 fps for two seconds and processed with the same methods as in the experiment \cite{Wang2017}. We see that the model provides a quantitative estimate of (i) the average edge-to-edge separation of the rafts, (ii) the increasing oscillations around the mean with decreasing frequency, and (iii) the critical collapse frequency. This agreement confirms that our model has  identified the key physical features of the problem and that it can be used to understand much of the dynamics of a pair of rafts. 

The behaviour not captured by our model is likely due to  the far-field hydrodynamic assumption and   the two-dimensional treatment of the system.  Two disks on an air water interface do not sit perfectly flat but rather tilt when in contact \cite{Vella2005} and the complex shape of the disks means that the magnetic dipole moments do not lie perfectly within the surface. When interacting with background fields, these three dimensional components can induce additional capillary interactions and modify the viscous drag, neither of which has been accounted for in this model.

Notably, if we apply this model to the $A=2~\mu$m raft experiments with the same set of parameters, the model captures the mean orbital position but significantly underestimates the value of the critical rotation frequency. The failure to capture the rotation frequency is again probably related to the various simplifications of the  model. In particular, three-dimensional effects are likely to play a larger role  here since   the raft amplitude increases and thus the two-dimensional assumptions breaks down.

\section{Conclusion}

Magnetic micro-disks floating on an air water interface and  with non-flat edge profiles
 exhibit both dynamic and static self-assembly. These 
  structures demonstrate a non-linear dependence on the shape of the disk and its magnetisation. In this paper we developed a series of simple models in order to illuminate the origin of this non-linear dependence and to provide a means to predict the resultant dynamics. 
  
  Modelling was achieved by breaking the behaviour into three physical components common to all experimental trajectories: (i) the static assembled configurations, (ii) the dynamics of the collapse, and (iii) the physics of the mean orbital position. When treated independently each of these phenomena can be explained through simple models of physical mechanics which consider the interplay of   surface energy, hydrodynamics and magnetic forces. Surprisingly, while only far-field interactions are needed to explain both the behaviour of the static assemblies and the dynamics of the collapse,   the mean orbital configurations is governed by the  capillary near field.

We next combined  
 these three separate models    in order to compare with the results from a series of new experiments employing micro-rafts with $\alpha=1/4$, $A=1~\mu$m for which the magnetic layer was applied over the course of 48 hours to reduce imperfections in the magnetic dipole moment. This two floating disk system was then driven by an external permanent magnet attached to a stepper motor. Assuming the magnetic parameters take the values of the previous experiments \cite{Wang2017}, the model successfully estimates the mean separation, the collapse location and the increasing oscillations seen by the experiments.

Through identifying the key physical features for   two interacting magneto-capillary disks, the current work provides a theoretical basis upon which  to develop further models to predict and control the dynamic and static programmability of multiple capillary rafts. Furthermore by extending the results to other edge undulations, this work could be used to explore how non-identical rafts would interact.

\section*{Acknowledgements}
This project has received funding from the European Research Council (ERC) under the European Union's Horizon 2020 research and innovation programme  (grant agreement 682754 to EL). MS and WW thank Max Planck Society for financial support. WW also thanks the Alexander von Humboldt Foundation for a postdoctoral fellowship and a travel grant. 

%%%END OF MAIN TEXT%%%


\balance


%%%REFERENCES%%%
\bibliography{library_JournalNameAbbreviated} %You need to replace "rsc" on this line with the name of your .bib file
\bibliographystyle{rsc} %the RSC's .bst file

\end{document}